\documentclass[a4paper,twoside,10pt]{letter}
\usepackage{graphicx,saj,multicol,subeqnarray}
\usepackage{ wasysym }
\usepackage{longtable}
\usepackage{ gensymb }

\newcommand{\HII}{H\,{\sc ii}}

\def\arcmin{\hbox{$^\prime$}}
\def\arcsec{\hbox{$^{\prime\prime}$}}

\def\p0{\phantom{0}}

\def\papertype{}

\def\udc{...}
\begin{document}
\baselineskip=3.1truemm
\columnsep=.5truecm
\newenvironment{lefteqnarray}{\arraycolsep=0pt\begin{eqnarray}}
{\end{eqnarray}\protect\aftergroup\ignorespaces}
\newenvironment{lefteqnarray*}{\arraycolsep=0pt\begin{eqnarray*}}
{\end{eqnarray*}\protect\aftergroup\ignorespaces}
\newenvironment{leftsubeqnarray}{\arraycolsep=0pt\begin{subeqnarray}}
{\end{subeqnarray}\protect\aftergroup\ignorespaces}
%

% Running titles

\markboth{\eightrm NEW 20-CM RADIO-CONTINUUM STUDY OF THE SMALL MAGELLANIC CLOUD: PART II - POINT SOURCES}
{\eightrm G. F. WONG, et. al.}

{\ }

\publ

\type

{\ }

% Title

\title{New 20-cm Radio-continuum study of the Small Magellanic Cloud: part II -- Point sources}

% Authors

\authors{G. F. Wong, M. D. Filipovi\'c, E. J. Crawford,  N.~F.~H. Tothill, A.~Y. De Horta }
\authors{D. Dra\v skovi\'c, T.~J. Galvin, J.~D. Collier,  J.~L. Payne }

\vskip3mm

% Address

\address{University of Western Sydney, Locked Bag 1797, Penrith South DC, NSW 2751, AUSTRALIA}
\Email{m.filipovic}{uws.edu.au}

% Received and Accepted dates

\dates{October XX, 2011}{October XX, 2011}

% Abstract
\summary{We present a new catalogue of radio-continuum sources in the field of the Small Magellanic Cloud (SMC). This catalogue contains sources previously not found in 2370~MHz ($\lambda$=13~cm) with sources found at 1400~MHz ($\lambda$=20~cm) and 843 MHz ($\lambda$=36~cm). 45 sources have been detected at 13~cm, with 1560 sources at 20~cm created from new high sensitivity and resolution radio-continuum images of the SMC at 20~cm from Paper I . We also created a 36~cm catalogue to which we listed 1689 radio-continuum sources. } 
% Keywords (see keywords.pdf file)


\begin{multicols}{2}
{

% Sections
%%%%%%%%%%%%%%%%%%%%%%%%%%%%%%%%%%%%%%%%%%%%
\section{1. INTRODUCTION}
%%%%%%%%%%%%%%%%%%%%%%%%%%%%%%%%%%%%%%%%%%%%

The Small Magellanic Cloud (SMC), with its well established distance ($\sim$60~kpc; Hilditch et al.~2005) and ideal location in the coldest areas of the radio sky towards the South Celestial Pole, allows observation of radio emissions to be made without interference from Galactic foreground radiation. This means that the SMC is an ideal location in which to study radio sources such as supernova remnants (SNRs; Filipovi{\' c} et al. 2005, 2008), \HII\ regions and Planetary Nebulae (PNe; Filipovi{\' c} et al. 2009a) which may be difficult to study in our own and other  more distant galaxies.  

Over the last 40 years extensive radio-continuum surveys of the SMC have been made including interferometric observations using the Molonglo Obervatory Synesis Telescope (MOST; Ye et al. 1995) and Australia Telescope Compact Array (ATCA; Filipovi{\' c} et al. 2002, Payne et al. 2004, Filipovi{\' c} et al. 2009b, Mao et al. 2008, Dickel et al. 2010), and single dish observations from the 64-m Parkes radio-telescope (Filipovi{\' c} et al. 1997, 1998).    

Catalogues of radio-continuum point sources towards the SMC have been produced from these surveys, and from wider surveys of the southern sky. The first SMC source catalogue was produced by McGee, Newton \& Butler (1976) using the Parkes radio telescope at 5009~MHz ($\lambda$=6~cm); it contained 27 sources, 13 of which were also detected at 8800~MHz ($\lambda$=3.4~cm). The resolution of the observations was limited to 4\arcmin\ at 6~cm and 2.7\arcmin\ at 3.4~cm. 
}
\end{multicols}

%%%%%%%%%%%% Table 1
\clearpage
\centerline{{\bf Table 1.} Summary of previous radio-continuum source catalogues of the SMC.}
\vskip1mm
\centerline{
\begin{tabular}{rccccc}
\hline
\emph{Telescope}&\emph{Freq}&\emph{Beam Size}&\emph{No of Sources} &\emph{Reference}\\
& (MHz) &(arcmin)& \emph{Detected}&&\\
\hline
Molonglo& \p0408 & 2.62 $\times$ 2.86 & \p075 & 1\\
Parkes &  1400  & 15.0 & \p021 & 2\\
Parkes &  2700  & 7.7 & \p025 & 3\\
Parkes &  5009 & 4.0 & \p027 & 2\\
Parkes &  8800  & 2.5 & \p013 & 2\\
Parkes & 1400 & 15.0 & \p028 & 4\\
MOST   &  \p0843 & 0.75 & ~450 & 5\\
Parkes &  1420  & 13.8 & \p085 & 6\\
Parkes &  2450  & 9.0 & 107 & 6\\
Parkes &  4750  & 4.5 & \p099 & 6\\
Parkes &  4850  & 4.9 & 187 & 6\\
Parkes &  8550  & 2.7 & \p041 & 6\\
ATCA   &  1420  & 1.63 & 534 & 7\\
ATCA   &  2370  & 0.67 & 697 & 7\\
ATCA   &  4800  & 0.5 & \p075 & 7\\
ATCA   &  8640  & 0.25 & \p054 & 7\\
\hline
\end{tabular}}
\centerline{1. Clarke et al. (1976), 2. McGee et al. (1976), 3. PKSCAT-90, 4. Haynes et al.(1986),}
\centerline{ 5. Turtle et al. (1998), 6. Filipovi\'c et al. (1997), 7. Filipovi\'c et al. (2002)}
\vskip.5cm
%%%%%%%%%%%%%%%%%%%%%%%%%%%%%%%%%%%%%%%%%

\begin{multicols}{2}
{
%%%%%%%%%%%%%%%%%%%%%%%%%%%%%%%%%%%%%%%%%
From the mid-1970s to the present, other surveys have been performed, increasing the number of sources detected (see Table~1). We recently published a set of new high-resolution radio-continuum mosaic images of the SMC at 1400~MHz ($\lambda$ = 20~cm), created by combining observations from ATCA and Parkes (Wong et al. 2011, hereafter Paper~I). 

We now present a catalogue of radio-continuum sources towards the SMC derived from an 2370~MHz ($\lambda$ = 13~cm) mosaic image from Filipovi\'c et al. (2002), one of our 20~cm mosaic radio-continuum images (Fig.~2 in Paper~I) and from an 843~MHz ($\lambda$ = 36~cm) MOST image (Turtle et al.~1998). In \S2 we describe the data used to derive the radio-continuum point sources. In \S3 we describe our source fitting and detection methods. \S4 contains our conclusions and the appendix contains the radio-continuum source catalogue. 

%%%%%%%%%%%%%%%%%%%%%%%%%%%%%%%%%%%%%%%%%%%%
\section{2. DATA}
%%%%%%%%%%%%%%%%%%%%%%%%%%%%%%%%%%%%%%%%%%%%

The 13~cm radio-continuum catalogue was produced from a SMC mosaic radio survey of 20 square degrees (Filipovi\'c et al. 2002). These observations have a beam size of $\sim$40\arcsec\ and r.m.s. noise of 0.4~mJy/beam.

The 20~cm mosaic image (Fig.~2 in Paper~I) was created by combining data from ATCA project C1288 (Mao et al.~2008)  with data obtained for a Parkes radio-continuum study of the SMC (Filipovi\'c et al. 1997). This image has a beam size of 17.8\arcsec $\times$ 12.2\arcsec\ with r.m.s. noise of 0.7~mJy/beam.\\\\\\\\

The 36~cm image comes from a radio survey of 36 square degrees containing the SMC (Turtle et al.~1998). These observations have a beam size of $\sim$45\arcsec\ and r.m.s. noise of 0.7~mJy/beam --- approximately equal to that of the 20~cm image.

Table~2 contains the field size of all the images used to derive the radio-continuum sources contained in this paper (Appendix).  \\\\

\centerline{{\bf Table 2.}  Field size of images used.}
\vskip1mm
\centerline{
\begin{tabular}{ccccccccc}
\hline
\emph{Image} &\emph{RA$_1$} &\emph{RA$_2$}&\emph{ DEC$_1$}& \emph{DEC$_2$}\\
\hline
13~cm&00\degree 27\arcmin&01\degree 35\arcmin&-70\degree 30\arcmin&-75\degree 15\arcmin&\\
20~cm&00\degree 10\arcmin&01\degree 43\arcmin&-69\degree 16\arcmin&-75\degree 40\arcmin&\\
36~cm&00\degree 16\arcmin&01\degree 40\arcmin&-72\degree 30\arcmin&-74\degree 38\arcmin&\\
\hline
\end{tabular}}
\vskip.5cm

%%%%%%%%%%%%%%%%%%%%%%%%%%%%%%%%%%%%%%%%%%%%
\section{3. SOURCE FITTING AND DETECTION}
%%%%%%%%%%%%%%%%%%%%%%%%%%%%%%%%%%%%%%%%%%%%

The \textsc{miriad} task \textsc{imsad} (Sault \& Killeen 2010) was used to detect sources in the 20~cm and 36~cm images, requiring a fitted Gaussian flux density $>$5$\sigma$ (3.5~mJy). All sources were then visually examined to confirm that they are genuine point sources, excluding extended emission, bright side lobes, etc.  

The radio-continuum sources catalogued in Table~A1, are extra sources at 13~cm that were not previously identified as part of the 13~cm catalogue taken from Filipovi\'c et al (2002).  The 13~cm catalogue retrieved from Filipovi\'c et al. (2002) was detected with a fitted Gaussian flux density of $>$5$\sigma$ (2.0~mJy).  Sources catalogued in Table~A1 were visually found with a $\sigma$ between 3$\sigma$ and 5$\sigma$.  \\\\

%%%%%%%%%%%%%%%%%%%%%%%%%%%%%%%%%%%%%%%%%%%%%%%%
}
\end{multicols}
%%%%%%%%%%%% Table 1%%%%%%%%%%%%%%%%%%%%%%%%%%%%%%%%

\centerline{{\bf Table 3.} Information of the images and catalogue of radio-continuum sources }
\vskip1mm
\centerline{
\begin{tabular}{ccccc}
\hline
\emph{$\lambda$ (cm)}&\emph{ RMS (mJy/beam)} &\emph{Number of Sources}&\emph{Within the Field of the 13~cm image} &\emph{Beam Size (arcsec)}\\
\hline
13~cm&0.4&743*&743*&45\\
20~cm&0.7&1560&824&14.8$\times$12.2\\
36~cm&0.7&1689&1198&40\\
\hline
\end{tabular}}
\centerline{* Values include the original catalogue retrieved from Filipovi\'c et al.~(2002)}
\vskip.5cm

%%%%%%%%%%%%%%%%%%%%%%%%%%%%%%%%%%%%%%%%%

\begin{multicols}{2}
{
%%%%%%%%%%%%%%%%%%%%%%%%%%%%%%%%%%%%%%%%%

The catalogue of radio-continuum sources contains positions RA(J2000), Dec(J2000) and integrated flux densities at 13~cm (Table~A1), 20~cm (Table~A2) and 36~cm (Table~A3).  Table~3 contains the r.m.s., number of sources detected, number of sources identified within the field of the 13~cm image and beam size for each image.   

%%%%%%%%%%%%%%%%%%%%%%%%%%%%%%%%%%%%%%%%%%%%
\section{4. CONCLUSION}
%%%%%%%%%%%%%%%%%%%%%%%%%%%%%%%%%%%%%%%%%%%%

We present a new catalogue of radio-continuum sources towards the SMC, containing sources previously not identified at 13~cm and sources found at 20~cm and 36~cm.  

The 13~cm catalogue contains 45 sources from a mosaic 13~cm radio survey (Table~A1; Filipovi\'c et al. 2002). Containing 1560 sources (Table~A2) the 20~cm catalogue, has been created from new high-sensitivity and resolution radio-continuum images of the SMC at 20 cm from Paper I . We also created a 36~cm catalogue to which we listed 1689 radio-continuum sources (Table~A3), created from a MOST radio survey of the SMC (Turtle et al.~1998).

%%%%%%%%%%%%%%%%%%%%%%%%%%%%%%%%%%%%%%%%%%%%
% Acknowledgements
%%%%%%%%%%%%%%%%%%%%%%%%%%%%%%%%%%%%%%%%%%%%

\acknowledgements{The Australia Telescope Compact Array and Parkes radio telescope  are parts of the Australia Telescope National Facility which is funded by the Commonwealth of Australia for operation as a National Facility managed by CSIRO. This paper includes archived data obtained through the Australia Telescope Online Archive (http://atoa.atnf.csiro.au). We used the {\sc karma} and {\sc miriad} software package developed by the ATNF. }

%%%%%%%%%%%%%%%%%%%%%%%%%%%%%%%%%%%%%%%%%%%%
% References
%%%%%%%%%%%%%%%%%%%%%%%%%%%%%%%%%%%%%%%%%%%%

\references

Clarke. J.N., Little A.G., Mills B.T.: 1979 \journal{Aust. J. Phys. Astrophys. Suppl } \vol{40}, 1.

Dickel, J.R.; Gruendl, R.A.; McIntyre, V.J., Shaun W.A.: 2010, \journal{Astron. J.}, \vol{140}, 1511.

Filipovi{\'c}, M.D., Jones, P.A., White, G.L, Haynes, R.F, Klein, U., Wielebinski, R.: 1997, \journal{Astron. Astrophys. Suppl. Series}, \vol{121}, 321.

Filipovi{\'c}, M.D., Haynes, R.F., White, G.L., Jones, P.A.: 1998, \journal{Astron. Astrophys. Suppl. Series}, \vol{130}, 421.

Filipovi{\'c}, M.D., Bohlsen, T., Reid, W, Staveley-Smith, L., Jones, P.A, Nohejl, K., Goldstein, G.: 2002, \journal{Mon. Not. R. Astron. Soc.}, \vol{335}, 1085.

Filipovi{\'c}, M.D., Payne, J.L., Reid, W., Danforth, C.W., Staveley-Smith, L., Jones, P.A., White, G.L.: 2005, \journal{Mon. Not. R. Astron. Soc.}, \vol{364}, 217.

Filipovi{\'c}, M.D., Haberl, F., Winkler, P.F., Pietsch, W., Payne, J.L., Crawford, E.J., de Horta, A.Y., Stootman, F.H., Reaser, B.E.: 2008, \journal{Astron. Astrophys.}, \vol{485}, 63.

Filipovi{\'c}, M.D., Cohen, M., Reid, W.A., Payne, J.L., Parker, Q.A., Crawford, E.J., Boji\v ci\'c, I.S., de Horta, A.Y., Hughes, A., Dickel, J., Stootman, F.: 2009a, \journal{Mon. Not. R. Astron. Soc.}, \vol{399}, 769.

Filipovi{\'c}, M.D., Crawford E.~J., Hughes A., Leverenz H., de Horta A.~Y., Payne J.~L., Staveley-Smith L., Dickel J.~R., Stootman F.~H., White G.~L.: 2009b, in van Loon J.~T., Oliveira J.~M., eds, \journal{IAU Symposium Vol. 256 of IAU Symposium}, PDF8

Haynes, R. F., Murray, J. D., Klein, U.,  Wielebinski, R.:1986,  \journal{Astron. Astrophys.}, \vol{159}, 22.

Hilditch, R.W., Howarth, I.D., Harries, T.J.: 2005, \journal{Mon. Not. R. Astron. Soc.}, \vol{357}, 304.

Mao, S.A., Gaensler, B.M., Stanimirovi{\'c}, S., Haverkorn, M., McClure-Griffiths, N.M., Staveley-Smith, L., Dickey, J.M.: 2008, \journal{Astrophys. J.}, \vol{688}, 1029.

McGee, R. X., Newton, L. M., Butler, P. W.: 1976, \journal(Aust. J. of Phys), \vol(29), 329.

Payne, J.L., Filipovi{\'c}, M.D., Reid, W., Jones, P.A., Staveley-Smith, L., White, G.L.: 2004, \journal{Mon. Not. R. Astron. Soc.}, \vol{355}, 44.

Sault, R., Killeen, N.: 2010, Miriad Users Guide, ATNF

Turtle, A.J., Ye, T., Amy,~S.W., Nicholls,~J.: 1998, \journal{Publ. Astron. Soc. Aust}, \vol{15}, 280.

Wong, G.F., Filipovi{\'c}, M.D., Crawford, E.J., de Horta, A.Y., Galvin, T., Dra\v skovi{\'c}, D., Payne, J.L.: 2011, \journal{Serb. Astron. J.}, \vol{182}, 43.

Ye, T. S., Amy, S. W., Wang, Q. D., Ball, L., Dickel, J.: 1995, \journal{Mon. Not. R. Astron. Soc.}, \vol{275}, 1218.

\endreferences

}
\end{multicols}

\newpage
%%%%%%%%%%%%%%%%%%%%%%%%%%%%%%%%%%%%%%%%%%%%
\section{APPENDIX}
%%%%%%%%%%%%%%%%%%%%%%%%%%%%%%%%%%%%%%%%%%%%

%table
\centerline{{\bf Table A1.}  13~cm Catalogue of point sources in the field of the SMC with integrated flux density.}
\vskip1mm
\setlongtables
\begin{longtable}{cccccccc}
\hline
\emph{Name}&\emph{R.A.$_{2370}$}&\emph{Dec.$_{2370}$}&\emph{ S$_{13}$}\\
\emph{ATCA}&\emph{ (J2000)} &\emph{(J2000)}&\emph{(mJy)}\\
\hline
J003414-722655 & 00:34:14.1 & -72:26:55.5 & 2.51 \\
J003446-733321 & 00:34:46.0 & -73:33:21.0 & 2.77 \\
J003609-740501 & 00:36:09.8 & -74:05:01.2 & 2.79 \\
J003850-732947 & 00:38:50.4 & -73:29:47.4 & 2.57 \\\smallskip
J003903-735148 & 00:39:03.6 & -73:51:48.8 & 2.48 \\
J004021-735211 & 00:40:21.1 & -73:52:11.3 & 2.22 \\
\hline
\end{longtable}
\vskip.5cm

%%%%%%%%%%%%%%%%%%%%%%%%%%%%%%%%%%%%%%%%%%%%%%%%%%%%%%%%%%%
%table
\centerline{{\bf Table A2.}  20~cm Catalogue of point sources in the field of the SMC with integrated flux density.}
\vskip1mm
\setlongtables
\begin{longtable}{cccccccc}
\hline
\emph{Name}&\emph{R.A.$_{1400}$}&\emph{Dec.$_{1400}$}&\emph{ S$_{20}$}\\
\emph{ATCA}&\emph{ (J2000)} &\emph{(J2000)}&\emph{(mJy)}\\
\hline
J001013-732120 & 00:10:13.4 & -73:21:20.6 & 26.97 \\
J001014-753530 & 00:10:14.5 & -75:35:30.8 & 35.13 \\
J001055-724428 & 00:10:55.6 & -72:44:28.8 & 12.14 \\
J001008-732111 & 00:10:08.6 & -73:21:11.8 & 19.75 \\\smallskip
J001116-745723 & 00:11:16.2 & -74:57:23.1 & 14.11 \\
J001157-734955 & 00:11:57.4 & -73:49:55.4 & 107.45 \\
\hline
\end{longtable}
\vskip.5cm

%%%%%%%%%%%%%%%%%%%%%%%%%%%%%%%%%%%%%%%%%%%%%%%%%%%%%%%%%%%
%table
\centerline{{\bf Table A3.}   36~cm Catalogue of point sources in the field of the SMC with integrated flux density.}
\vskip1mm
\setlongtables
\begin{longtable}{ccccc}
\hline
\emph{Name}&\emph{R.A.$_{843}$}&\emph{Dec.$_{843}$}&\emph{ S$_{36}$}\\
\emph{ATCA}&\emph{ (J2000)} &\emph{(J2000)}&\emph{(mJy)}\\
\hline
J001459-750128 & 00:14:59.8 & -75:01:28.7 & 14.38 \\
J001526-744248 & 00:15:26.3 & -74:42:48.7 & 12.61 \\
J001544-744902 & 00:15:44.4 & -74:49:02.9 & 3.75 \\
J001557-744558 & 00:15:57.8 & -74:45:58.1 & 10.23 \\\smallskip
J001601-741658 & 00:16:01.2 & -74:16:58.5 & 6.92 \\
J001614-741446 & 00:16:14.0 & -74:14:46.5 & 3.28 \\
\hline
\end{longtable}
\vskip.5cm

%********** one column figure *************************

\vskip.5cm
%%%%%%%%%%%%%%%%%%%%%%%%%%%%%%%%%%%%%%%%%

\vfill\eject

{\ }

% Serbian abstract

% Title

\naslov{NOVO PROUQAVA{NJ}E MALOG MAGELANOVOG OBLAKA U RADIO-KONTINUMU NA 20~CM: DEO~{\bf II} - SNIMCI}

% Authors

\authors{G. F. Wong, M. D. Filipovi\'c, E. J. Crawford,  N.~F.~H. Tothill, A.~Y. De Horta,} 
\authors{ D. Dra\v skovi\'c, T.~J. Galvin, J.~D. Collier , J. L. Payne }

\vskip3mm

% Address

\address{University of Western Sydney, Locked Bag 1797, Penrith South DC, NSW 2751, AUSTRALIA}

\Email{m.filipovic}{uws.edu.au}

\vskip3mm

% UDC

\centerline{\rrm UDK \udc}

\vskip1mm

\centerline{\rit Originalni nauqni rad}

\vskip.7cm

\begin{multicols}{2}

{

% Abstract

\rrm 

We present a new catalogue of radio-continuum sources in the field of the Small Magellanic Cloud (SMC). This catalogue contains sources previously not found in 2370~MHz ($\lambda$=13~cm) with sources found at 1400~MHz ($\lambda$=20~cm) and 843 MHz (36~cm). 46 sources have been detected at 13~cm, with 1560 sources at 20~cm created from new high sensitivity and resolution radio-continuum images of the SMC at 20~cm from Paper I . We also created a 36~cm catalogue to which we listed 1689 radio-continuum sources
}

\end{multicols}

\end{document}